\documentclass[twocolumn,showpacs,amsmath,amssymb,aps,prd,reprint]{revtex4-1}

\usepackage[latin1]{inputenc}
\usepackage{graphicx} 
\usepackage{dcolumn}
\usepackage{bm}
\usepackage{ifpdf}

\begin{document}

\title{Palatini $f(R)$ Black Holes in Nonlinear Electrodynamics}

\author{Gonzalo J. Olmo$^1$} \email{gonzalo.olmo@csic.es}
\author{D. Rubiera-Garcia$^2$} \email{rubieradiego@gmail.com}

\affiliation{$^1$Departamento de F\'{i}sica Te\'{o}rica and IFIC, Centro Mixto Universidad de
Valencia - CSIC. Universidad de Valencia, Burjassot-46100, Valencia, Spain}
\affiliation{$^2$Departamento de F\'{i}sica, Universidad de Oviedo, Avenida Calvo Sotelo 18, 33007, Oviedo, Asturias, Spain}


\date{\today}

\begin{abstract}
The electrically charged Born-Infeld black holes in the Palatini formalism for $f(R)$ theories are analyzed. Specifically we study those supported by a theory $f(R)=R\pm R^2/R_P$, where $R_P$ is Planck's curvature. These black holes only differ from their General Relativity counterparts very close to the center, but may give rise to different geometrical structures in terms of inner horizons. The nature and strength of the central singularities are also significantly affected. In particular, for the model $f(R)=R - R^2/R_P$ the singularity is shifted to a finite radius, $r_+$, and the Kretschmann scalar diverges only as $1/(r-r_+)^{2}$ . \end{abstract}
\maketitle

\section{Introduction}

Black holes are one of the most intriguing objects of Nature.
Their event horizons dramatically modify the causal structure
of space-time generating regions from where nothing can scape
classically. From a semiclassical perspective, however, event
horizons are responsible for the generation of quantum radiation
that slowly forces their evaporation in a manner already described
by Hawking almost 40 years ago \cite{Hawking-1975,FNS-2005,Parker-Toms}. In terms of the Planck length,
 $l_P\equiv \sqrt{\hbar G/c^3} \approx 10^{-35}$m, the event horizon of macroscopic
black holes is located very far away from the center, where a
singularity is expected classically. This singularity, however,
should be an artifact of the classical theory, and a full quantum theory
of gravity should provide suitable mechanisms to avoid its formation.
Within such a theory, the weak, perturbative quantum
phenomenon of Hawking radiation should be consistent with the
strongly non-perturbative quantum processes expected to happen at the
black hole center to remove the classically-expected singularity.
Therefore, in the vicinity of the singularity, the internal structure of black holes must be
different from that predicted by classical general relativity (GR). 

In this paper we address the issue of the internal structure of black
holes from a purely classical perspective by considering an $f(R)$
gravitational action containing a correction term that generates
non-perturbative effects at the Planck scale. This is achieved by
formulating the theory à la Palatini  \cite{Olmo-Review}, i.e., by considering metric
and connection as independent fields. As a result, the
dynamical degrees of freedom of the theory are the same as in GR,
which implies that there is no multiplicity of solutions. In fact,
the number of solutions is the same as in GR, and their structure
is significantly deformed near (and only near) the
characteristic scale set by the correcting terms of the action. This
means that the internal structure of these black holes is different
from that of GR only near the Planck scale, while for lengths much
larger than $l_P$ the departures from GR are totally negligible.
We hope that the analysis of this type of solutions allows to
provide new insights on how quantum gravitational effects could
affect the interior of black holes.

The modified dynamics generated by the Palatini version of $f(R)$
theories is intrinsically different from that of their metric counterpart,
where the connection is assumed to be defined in terms of the Christoffel
symbols of the metric. In the usual (metric) formulation, the modified
dynamics is due to the existence of an effective dynamical scalar degree of freedom
that can be defined as $\phi=df/dR$. In the Palatini approach, though a scalar-tensor representation is also possible, the resulting scalar is non-dynamical and, therefore, there are no new degrees of freedom \cite{Olmo:2011fh}. The equation for the
independent connection can be solved by introducing an auxiliary metric
that is conformally related with the physical metric $g_{\mu\nu}$. As we
will show later, the conformal factor turns out to be a function of the trace,
$T=g^{\mu\nu}T_{\mu\nu}$, of the energy-momentum tensor $T_{\mu\nu}$ of
the matter fields. As a result, the field equations for the metric contain a number of new terms that explicitly depend on $T$ and its derivatives.
When $T=0$, the field equations boil down to those of GR (with possibly an effective
cosmological constant, depending on the choice of $f(R)$ Lagrangian), but when
$T\neq 0$, the connection-induced matter terms yield modified dynamics \cite{Olmo-Review}.
According to this, Palatini $f(R)$ theories have the same (classical) vacuum solutions
as GR with a cosmological constant. However, if semiclassical effects
are taken into account (Hawking radiation, trace anomaly, \ldots) then black
hole solutions should be different for different $f(R)$ Lagrangians because then
the trace $T$ would no longer be zero. 

Besides semiclassical scenarios, which are difficult to treat consistently,
non-vacuum spacetimes that are of
interest to explore Planck-scale effects of the Palatini dynamics are those of
electrically charged black holes. Though the stress-energy tensor of Maxwell
electrodynamics is traceless and, therefore, cannot be used to probe the Palatini
dynamics, it seems reasonable to expect that besides the gravitational field, other
fields could develop non-linear corrections at high energy scales. Such is the case of nonlinear electrodynamics (NED), which are expected to arise as modifications of Maxwell theory in several scenarios, such as effective Lagrangians in Quantum Electrodynamics \cite{Dobado97}. Moreover, in order to break the tracelessness of the energy-momentum tensor and obtain electrically charged non-standard black holes NEDs is the natural choice. In this sense a physically well motivated candidate to be studied is Born-Infeld theory. The Born-Infeld (BI) Lagrangian was introduced in the thirties \cite{BI} in order to remove the divergence of the electron's self-energy in Classical Electrodynamics. Moreover, BI theory is singled out among the class of NEDs by its special properties concerning wave propagation such as the absence of shock waves and birefringence phenomena \cite{wave-BI}, and in addition it enjoys an electric-magnetic duality \cite{duality-BI}. In modern times, BI theory has received special attention as it arises in the low-energy regime of string and D-Brane physics \cite{string-BI}.

Electrically charged BI black hole solutions within the context of asymptotically flat Einstein gravity has been thoroughly studied and characterized for many years (see e.g. \cite{BI-gravity}). Other examples of gravitating non-linear electrodynamics have been considered \cite{BI-NED}, and moreover a general analysis on large families of gravitating NEDs has been also carried out \cite{BI-general}. Extensions of BI black holes to asymptotically (Anti-) de Sitter spaces can be found in Ref.\cite{BI-AdS}. Let us finally mention that NEDs have also been used as attempts to classically remove the black hole curvature singularity \cite{NED-regular}.

Beyond GR, black hole solutions within the context of higher order gravity theories (e.g. Lovelock theory \cite{Lovelock}) have attracted much attention in the last years, as these theories are suggested to arise as corrections to the Einstein-Hilbert action by some low-energy results of string theory \cite{ST}. In this context some authors have looked for black holes in Gauss-Bonnet theory coupled to BI electrodynamics  \cite{BI-GB}, as well as to related theories \cite{Aiello05}, finding large differences in terms of horizons and singularities as compared to their GR counterparts as well as striking new features such as the existence of branch singularities \cite{Wiltshire88}.

Literature on black hole solutions in the Palatini approach is rather scarce and has focused essentially on vacuum solutions of $f(R)$ theories. Most research at this regard deals with thermodynamics issues (see e.g. \cite{Bamba10} and references therein) and entropy definition \cite{entropy}. The aim of this paper is to provide an analysis on the geometrical structure of electrically charged black holes, through NEDs, and to compare it to the GR limit. In particular, we look for the deviance from GR of the structure of the BI black holes near their center, where the effects introduced by the Planck-scale corrected model $f(R)=R \pm R^2/R_P$ become relevant (throughout the paper $R_P\equiv l_P^{-2}$). These models have been studied previously in the cosmological context, where it has been shown that they can avoid the big bang singularity in isotropic and homogeneous scenarios \cite{BOSA-2009,BO-2010,MyTalks}, and also in astrophysical \cite{polytropes}, laboratory \cite{Olmo-2008a}, and solar system settings \cite{Olmo2005}. We provide a thorough analysis and discussion of the metric components, Kretschmann scalar, and inner horizons and characterize them according to the values of the relevant length scales involved. We find that, in general, the metric components can be solved in terms of power series expansions, which in some particular cases can be explicitly written in terms of special functions. Though the central singularity still appears in these models, we find that for the case $f(R)=R-R^2/R_P$ it is shifted to a finite radius, which we denote $r_+$, and its strength as measured by the Kretschmann scalar is significantly softened as compared to GR or the model $f(R)=R+R^2/R_P$, for which it occurs at $r=0$.

The paper is organized as follows. In section II we introduce the general setup and basic elements for both the modified gravity $f(R)$ and matter NED source, and solve the Einstein equations in Palatini formalism. In section III we introduce the BI model, that shall be used throughout the paper. Section IV is devoted to the study of black holes within $f(R)=R+R^2/R_P$ gravity theory coupled to BI NED, through the analysis of the Kretschmann scalar and the existence and features of the inner horizons. Then we extend this analysis to the $f(R)=R-R^2/R_P$ theory in section V and we conclude in section VI with a summary and some future perspectives.

\section{Definitions and Field Equations}

The action for GR coupled to Maxwell field is
\begin{equation}\label{eq:GR-ED}
S=\frac{1}{2\kappa^2}\int d^4x\sqrt{-g} R-\frac{1}{16\pi}\int d^4x \sqrt{-g}F_{\alpha\beta}F^{\alpha\beta} \ ,
\end{equation}
where $\kappa^2=8\pi G/c^3$, $R\equiv g^{\mu\nu}R_{\mu\nu}(\Gamma)$,  $R_{\mu\nu}(\Gamma)\equiv{R^\rho}_{\mu\rho\nu}$, ${R^\alpha}_{\beta\mu\nu}=\partial_\mu\Gamma_{\nu\beta}^\alpha-\partial_\nu\Gamma_{\mu\beta}^\alpha+\Gamma_{\mu\lambda}^\alpha\Gamma_{\nu\beta}^\lambda-\Gamma_{\nu\lambda}^\alpha\Gamma_{\mu\beta}^\lambda$ represents the components of the Riemann tensor, the field strength of the connection $\Gamma^\alpha_{\mu\beta}$, and $F_{\mu\nu}=\partial_{\mu}A_{\nu}-\partial_{\nu}A_{\mu}$ is the field strength tensor of the electromagnetic vector potential $A_\mu$. This definition leads to
\begin{equation}
T_{\mu\nu}\equiv -\frac{2}{\sqrt{-g}}\frac{\delta S_m}{\delta g^{\mu\nu}}=\frac{1}{4\pi}\left[{F_{\mu\alpha}} {F_\nu}^\alpha-\frac{F_{\alpha\beta}F^{\alpha\beta}}{4}g_{\mu\nu}\right] \ .
\end{equation}
Introducing the gauge field invariants as $X\equiv -\frac{1}{2}F_{\alpha\beta}F^{\alpha\beta}=\vec{E}^2-\vec{B}^2$ and $Y\equiv -\frac{1}{2}F_{\alpha\beta}F^{*\alpha\beta}=2\vec{E} \cdot \vec{B}$, with $F^{*\mu\nu}=\frac{1}{2}\epsilon^{\mu\nu\alpha\beta}F_{\alpha\beta}$ being the dual of the field strength tensor and $\vec{E}$ and $\vec{B}$ the electric and magnetic fields, respectively, the extension of (\ref{eq:GR-ED}) to the $f(R)$ case with NED source is just
\begin{equation}\label{eq:f-NED}
S=\frac{1}{2\kappa^2}\int d^4x\sqrt{-g} f(R)+\frac{1}{8\pi}\int d^4x \sqrt{-g}\varphi(X,Y) \ .
\end{equation}
where $\varphi(X,Y)$ is a given function of the two field invariants, and defines the particular NED model. Note that the condition $\varphi(X,Y)=\varphi(X-Y)$ for the NED Lagrangian density must be satisfied in order to implement parity invariance. Variation of the action (\ref{eq:f-NED}) with respect to the metric for a radial electric field ($E(r)\frac{\vec{r}}{r} \neq 0, \vec{B} =0$), which implies that $Y=0$, yields
\begin{equation}\label{eq:fR-eom}
f_R R_{\mu\nu}(\Gamma)-\frac{f}{2}g_{\mu\nu}=\frac{\kappa^2}{4\pi}\left[\varphi_X{F_{\mu\alpha}} {F_\nu}^\alpha+\frac{\varphi}{2}g_{\mu\nu}\right] \ .
\end{equation}
where we have introduced the notation  $f_R\equiv df/dR$ and $\varphi_X\equiv \frac{\partial \varphi}{\partial X}$. Taking the trace of (\ref{eq:fR-eom}) with the metric $g^{\mu\nu}$, we find that the scalar $R$ is algebraically related to the energy-momentum tensor through
\begin{equation}\label{eq:trace-P}
Rf_R-2f=\kappa ^2T \ .
\end{equation}
This algebraic equation generalizes the GR relation $R=-\kappa^2 T$ to non-linear Lagrangians and its solutions will be denoted by $R=R(T)$. 

The variation of (\ref{eq:f-NED}) with respect to $\Gamma^\lambda _{\mu \nu }$ must vanish
independently of (\ref{eq:fR-eom}) and gives (we assume a torsionless connection for simplicity \cite{Olmo-Review})
\begin{equation}\label{eq:field-G-simpler}
\nabla_\lambda\left(\sqrt{-g}f_R g^{\mu\nu}\right)=0 \ ,
\end{equation}
where $f_R\equiv f_R(R[T])$ must be seen as a function of the matter by virtue of (\ref{eq:trace-P}). This equation can be solved easily  (see for instance \cite{OSAT09}) and leads to
\begin{equation}\label{eq:Gamma-1}
\Gamma^\lambda_{\mu \nu }=\frac{h^{\lambda \rho
}}{2}\left(\partial_\mu h_{\rho \nu }+\partial_\nu
h_{\rho \mu }-\partial_\rho h_{\mu \nu }\right),
\end{equation}
where  $h_{\mu \nu }\equiv f_R g_{\mu \nu }$, i.e., $\Gamma^\lambda_{\mu \nu }$ is the Levi-Civita connection of a metric $h_{\mu\nu}$ conformally related with $g_{\mu \nu }$, being the conformal factor the function $f_R\equiv f_R(R[T])$. Using this expression for the connection, we can write (\ref{eq:fR-eom}) as an Einstein-like equation for the physical metric $g_{\mu\nu}$
\begin{eqnarray}\label{eq:Gmn}
G_{\mu \nu }(g)&=&\frac{\kappa
^2}{f_R}T_{\mu \nu }-\frac{R f_R-f}{2f_R}g_{\mu \nu
}\nonumber\\&-&\frac{3}{2(f_R)^2}\left[\partial_\mu f_R\partial_\nu
f_R-\frac{1}{2}g_{\mu \nu }(\partial f_R)^2\right]\nonumber \\
&+& \frac{1}{f_R}\left[\nabla_\mu \nabla_\nu f_R-g_{\mu \nu }\Box
f_R\right] \ .
\end{eqnarray}
One should note that all the $R, f,$ and $f_R$ terms on the right hand side of this system of equations are functions of the trace $T$ of the matter energy-momentum tensor. According to this, if $T=0$, those equations boil down to
\begin{equation}\label{eq:Gmn-vac}
G_{\mu \nu }(g)=\frac{\kappa
^2}{f_{R_0}}T_{\mu \nu }-\Lambda_{eff}g_{\mu\nu}
\end{equation}
where $\Lambda_{eff}\equiv (R_0 f_{R_0}-f_0)/2f_{R_0}$ is evaluated at the constant value $R_0=R[T=0]$ and plays the role of an effective cosmological constant. This observation highlights the fact that the modified dynamics of Palatini $f(R)$ theories can only arise if $T\neq 0$. For $T=0$, the theory becomes equivalent to GR$+\Lambda$ up to a constant rescaling of units.

The electromagnetic field satisfies the equation $\nabla_\mu \left(\varphi_X F^{\mu\nu}+ \varphi_Y F^{* \mu\nu}\right )=0$. For a diagonal metric with spherical symmetry and a purely radial electric field, for which only $F^{tr}$ is not zero, this equation becomes
\begin{equation}\label{eq:ned-eom}
\partial_r\left(\sqrt{-g_{tt}g_{rr}}r^2 \varphi_X F^{tr}\right)=0 \ \rightarrow \ \varphi_X F^{tr}=\frac{q}{r^2}\frac{1}{\sqrt{-g_{tt}g_{rr}}} \ .
\end{equation}
where $q$ is an integration constant identified as the electric charge for a given model. From this expression for $F^{tr}$ and the fact that $X=-g_{tt}g_{rr}(F^{tr})^2$, it follows that for any spherically symmetric metric
\begin{equation}\label{eq:Xvarphi}
\varphi_X^2 X=\frac{q^2}{r^4} \ .
\end{equation}
With this equation once a $\varphi(X,Y=0)$ theory is provided, the solution for $X=X(r)$ can be found, in principle, algebraically.

To continue with our analysis, it is convenient to express the NED stress-energy tensor as
\begin{equation}
{T_\mu}^\nu= -\frac{1}{4\pi}\left[\varphi_X{F_\mu}^\alpha {F_\alpha}^\nu-\frac{\varphi}{2}{\delta_\mu}^\nu\right] \ .
\end{equation}
Using matrix notation, with $\hat{I}$ representing the $2\times 2$ identity matrix and $\hat{0}$ a $2\times 2$ zero matrix, we find that
\begin{equation}
{F_\mu}^\alpha {F_\alpha}^\nu=X\begin{pmatrix}
 \hat{I} & \hat{0}  \\
\hat{0} & \hat{0}
\end{pmatrix} \ ,
\end{equation}
which can be used to write ${T_\mu}^\nu$ as
\begin{equation} \label{tracematrix}
{T_\mu}^\nu= \frac{1}{4\pi}\begin{pmatrix}
 \left(\frac{\varphi}{2}-X\varphi_X\right)\hat{I} & \hat{0}  \\
\hat{0} & \frac{\varphi}{2}\hat{I}
\end{pmatrix} \ .
\end{equation}
In order to guarantee the positive definiteness of the energy density for any NED model the condition
\begin{equation} \label{en-den}
\rho=T_t^t=\frac{1}{8\pi}\left(2\varphi_X \vec{E}^2-\varphi(X,Y=0)\right)\geq 0,
\end{equation}
will be assumed in what follows. This positive energy condition guarantees the single branched and monotonically decreasing character of the associated $E(r,q)$ field, as a consequence of Eq.(\ref{eq:Xvarphi}) (see Ref.\cite{BI-general}).

It is easy to see that the trace of the object (\ref{tracematrix}) leads to
\begin{equation}\label{eq:T-NED}
T=\frac{1}{2\pi}\left[\varphi-X\varphi_X\right]\ ,
\end{equation}
which vanishes for Maxwell electrodynamics $\varphi(X)=X$. This observation is very important because, as mentioned above, the modified dynamics of Palatini $f(R)$ theories can only arise if the trace $T\neq 0$.  Maxwell's linear electrodynamics times a constant is the only one (for vanishing $Y$) satisfying the $T=0$ condition. Any nonlinear generalization circumvents the tracelessness condition of the energy-momentum tensor, thus providing a deviance within Palatini $f(R)$ theories from their GR counterparts, as we shall see at once. Note that when the $Y$ invariant is considered, the above trace expression is extended as
\begin{equation}
T=\frac{1}{2\pi}[\varphi-X\varphi_X-Y\varphi_Y].
\end{equation}
The set of families with vanishing trace in this case is formed by all conic surfaces in the ($X,Y,\varphi$) plane having the origin as a vertex; for example the set of planes of the form $\varphi(X,Y)=aX+bY$ ($a,b$ constants) and all of them lead to the same solutions as for the GR case. However, here we shall be only interested in the purely electric case.

A glance at Eqs.(\ref{tracematrix}) and (\ref{eq:Xvarphi}) indicates that for a given NED the dependence of ${T_\mu}^\nu$ on the radial coordinate $r$ can be completely known before solving explicitly the equations for the metric. With the knowledge of ${T_\mu}^\nu$ one can proceed to solve for $g_{\mu\nu}$. To do it, one may consider directly (\ref{eq:Gmn}), whose right hand side involves several derivatives of $T$, or may consider the equations for $h_{\mu\nu}$ given in (\ref{eq:fR-eom}) and then use the conformal transformation to obtain $g_{\mu\nu}$. For notational and technical simplicity, we choose the second option. We begin by writing (\ref{eq:fR-eom}) as
\begin{equation}
{R_\mu}^\nu(h)=\frac{1}{f_R^2}\left(\kappa^2 {T_\mu}^\nu+\frac{f}{2}{\delta_\mu}^\nu\right) \ ,
\end{equation}
which in matrix notation becomes
\begin{equation}
{R_\mu}^\nu(h)=\frac{1}{f_R^2}\begin{pmatrix}
 \left[\frac{f}{2}+\frac{\kappa^2}{4\pi}\left(\frac{\varphi}{2}-X\varphi_X\right)\right]\hat{I} & \hat{0}  \\
\hat{0} & \left[\frac{f}{2}+\frac{\kappa^2}{8\pi}{\varphi}\right]\hat{I}
\end{pmatrix} \ .
\end{equation}
We now need to find the differential equations for the metric $h_{\mu\nu}$ and use the conformal relation between metrics to obtain $g_{\mu\nu}$. To simplify the field equations for $h_{\mu\nu}$ we define the line element in the two frames as follows
\begin{eqnarray}\label{eq:g-vs-h}
ds^2&=&-g_{tt}dt^2+g_{rr}dr^2+r^2d\Omega^2=\frac{1}{f_R}d\tilde{s}^2\nonumber \\ &=&\frac{1}{f_R}\left(-A(\tilde{r})e^{\psi(\tilde{r})}dt^2+\frac{d\tilde{r}^2}{A(\tilde{r})}+\tilde{r}^2d\Omega^2\right).
\end{eqnarray}
Therefore, to solve for the metric $g_{\mu\nu}$ we will solve first for $h_{\mu\nu}$ and then use the relation (\ref{eq:g-vs-h}) to find $g_{\mu\nu}$. Note in this sense that $r$ and $\tilde{r}$ are related by $\tilde{r}^2 =r^2f_R$.

With this decomposition of $h_{\mu\nu}$ and that choice of coordinate $\tilde{r}$, we can use the expressions for the components of ${R_\mu}^\nu(h)$ to obtain the following expressions
\begin{eqnarray}\label{eq:psi}
{R_t}^t-{R_{\tilde{r}}}^{\tilde{r}}&\equiv& -\frac{2}{h_{\tilde{r}\tilde{r}}}\frac{\psi_{\tilde{r}}}{\tilde{r}}=0 \\
{R_\theta}^\theta&\equiv& \frac{1}{\tilde{r}^2}\frac{d}{d\tilde{r}}\left[\tilde{r}(1-A(\tilde{r}))\right]=\frac{f+\frac{\kappa^2}{4\pi}\varphi}{2f_R^2}.
\end{eqnarray}
The first of these equations implies that $\psi(\tilde{r})$ is a constant, which can be absorbed into a redefinition of the time coordinate and will thus be omitted. On the other hand, taking into account the relation $\tilde{r}^2 =r^2f_R$ and defining $A(\tilde{r})=1-2M(\tilde{r})/\tilde{r}$, we find that
\begin{equation}
M_{\tilde{r}}=\frac{\left(f+\frac{\kappa^2}{4\pi}\varphi\right)r^2}{4f_R} \ ,
\end{equation}
which in terms of the $r$ variable becomes
\begin{equation}\label{eq:Mr}
M_r=\frac{\left(f+\frac{\kappa^2}{4\pi}\varphi\right)r^2}{4f_R^{3/2}}\left(f_R+\frac{r}{2}f_{R,r}\right) \,
\end{equation}
providing the right expression in the GR limit ($f_R \rightarrow 1$)
\begin{equation} \label{eq:GR-M}
M_r=\frac{\kappa^2}{8\pi}\left[X\varphi_X-\frac{\varphi}{2}\right]=-\frac{\kappa^2}{2}r^2 T_t^t.
\end{equation}
The equation (\ref{eq:Mr}) is far too complicated to be solved for a generic $f(R)$ gravity and $\varphi(X,Y=0)$ NED model. Hence we are led to consider particular cases of theories for both the matter and gravity fields, as described in the following sections.

\section{Born-Infeld NED}

In this case, the NED Lagrangian is defined as
\begin{equation}
\varphi(X)=2\beta^2\left(1-\sqrt{1-\frac{X}{\beta^2}-\frac{Y^2}{4\beta^4}}\right) \ .
\end{equation}
For this theory and considering electrically charged solutions, Eq.(\ref{eq:Xvarphi}) becomes
\begin{equation}
\frac{X}{1-\frac{X}{\beta^2}}=\frac{q^2}{r^4} \ ,
\end{equation}
which leads to
\begin{equation}
X=\frac{q^2 \beta ^2}{q^2+r^4 \beta ^2} \ .
\end{equation}
Since we will be dealing with expressions of the form $X/\beta^2$, it seems convenient to write that expression as follows
\begin{equation}
\frac{X}{\beta ^2}=\frac{q^2 }{q^2+\beta ^2 r^4}=\frac{1}{1+\frac{\beta^2r^4}{q^2}}=\frac{1}{1+z^4} \ ,
\end{equation}
where $r^4= \frac{q^2z^4}{\beta^2}$. With this notation we can write
\begin{eqnarray}
\varphi&=&2\beta^2\left(1-\frac{1}{\sqrt{1+\frac{1}{z^4}}}\right)\\
\varphi_X&=& \sqrt{1+\frac{1}{z^4}} \ ,
\end{eqnarray}
where the limit to the usual electrodynamics corresponds to $z\gg 1$.

For further reference we also need the expression for the BI $T_t^t$ component
\begin{equation} \label{tensorBI}
T_t^t=\frac{\beta^2}{4\pi} \left(\frac{\sqrt{z^4+1}}{z^2}-1\right)
\end{equation}
which obviously satisfies the positive energy condition (\ref{en-den}) for any $\beta$. The energy associated to the ESS field is given by

\begin{eqnarray}\label{energy}
\varepsilon(q)&=& 4\pi \int_0^{\infty}dr r^2 T_t^t(r,q)= \\ &=&
\beta^{1/2}q^{3/2} \int_0^{\infty} dt (\sqrt{t^4+1}-t^2) = \frac{\pi^{3/2} \beta^{1/2}}{3\Gamma(3/4)^2} q^{3/2}. \nonumber
\end{eqnarray}
Indeed this quantity plays a relevant role in the characterization of the GR BI black holes \cite{BI-gravity}. Note that using the scaling with $q$ in Eq.(\ref{energy}), the quantity $\frac{\varepsilon(q=1)}{\beta^{1/2}}=\frac{\pi^{3/2}}{3 \Gamma(3/4)^2} \simeq 1.236$ becomes a universal constant for a given parameter $\beta$ in the BI model.

\section{Study of $f(R)=R+R^2/R_P$}

We now consider the quadratic model $f(R)=R+R^2/R_P$, for which $R=-\kappa^2T$, as follows from Eq.(\ref{eq:trace-P}). From the general expression (\ref{eq:T-NED}) we find that
\begin{equation}
T=-\frac{\beta^2}{2\pi\sqrt{1+\frac{1}{z^4}}}\left[\frac{1}{z^4}+2\left(1-\sqrt{1+\frac{1}{z^4}}\right)\right] \ ,
\end{equation}
which can be expanded for large $z$ as $T/\beta^2 \approx -\frac{1}{8 \pi  z^8}+O\left[\frac{1}{z}\right]^{12}$, and for $z\to 0$ as $T/\beta^2 \approx -\frac{1}{\pi  z^2}+\frac{1}{\pi }-\frac{3 z^2}{4 \pi }+\frac{5 z^6}{16 \pi }+O[z]^7$. This behavior at small $z$ indicates that $f_R=1-2\kappa^2T/R_P$ may vanish at some small $z$ depending on the sign and magnitude of the combination $\lambda\equiv\kappa^2\beta^2/R_P$. In particular, for $f(R)=R+R^2/R_P$, $f_R$ does not vanish anywhere, whereas for $f(R)=R-R^2/R_P$ the function $f_R$ vanishes at a finite radius, which requires an independent analysis (see Section \ref{R-R2}). 

In terms of the variable $z$, (\ref{eq:Mr}) can be written as
\begin{equation}\label{eq:Mz}
M_z=\frac{\gamma^3z^2}{4f_R^{3/2}}\left(f+\frac{\kappa^2}{4\pi}\varphi\right)\left(f_R+\frac{z}{2}f_{R,z}\right) \ ,
\end{equation}
where $\gamma\equiv \sqrt{q/\beta}$. Now we just need to integrate this function numerically in each case. Additionally, it is useful to note that the dimensions of the term $\left(f+\frac{\kappa^2}{4\pi}\varphi\right)$ are given by a factor $\kappa^2\beta^2$, in other words, $\left(f+\frac{\kappa^2}{4\pi}\varphi\right)=\kappa^2\beta^2\left(\tilde{f}+\frac{1}{4\pi}\tilde{\varphi}\right)$, where $\tilde{f}=-\tilde{T}+\frac{\kappa^2\beta^2}{R_P}\tilde{T}^2$, $\tilde{T}=T/\beta^2$, and  $\tilde{\varphi}=\varphi/\beta^2$. We thus find that the analysis becomes more transparent if we consider the dimensionless variable $\tilde{M}\equiv M/(\gamma^3\kappa^2\beta^2)$, which leads to
\begin{equation}\label{eq:Mzt}
\tilde{M}_z=\frac{z^2}{4f_R^{3/2}}\left(\tilde{f}+\frac{\tilde{\varphi}}{4\pi}\right)\left(f_R+\frac{z}{2}f_{R,z}\right) \ .
\end{equation}
With this choice of variables, the only adjustable parameter is $\lambda\equiv \frac{\kappa^2\beta^2}{R_P}$, which appears in $\tilde{f}$ and $f_R$.

\subsection{External horizon and the function $G(z)$}

Horizons arise when $g_{tt}=0$. According to the decomposition of the metric that we have considered, this implies that
$g_{tt}=-A(\tilde{r})/f_R=0$, which is equivalent to $A(\tilde{r})=0$ (and also to $f_R\to \infty$, which occurs at $z=0$). According to our analysis, we have
\begin{equation}
A(\tilde{r})=1-\frac{2M(\tilde{r})}{\tilde{r}} \ ,
\end{equation}
where $M(\tilde{r})$ can be expressed as
\begin{eqnarray}
\hat{M}(z)\equiv\frac{M(z)}{M_0}&=&1+\frac{\gamma^3\kappa^2\beta^2}{M_0} G(z)\\
G(z)&=& -\int_z^\infty dz' \tilde{M}_{z'} \ ,
\end{eqnarray}
where $M_0$ is an integration constant that represents the Schwarzschild mass in the vacuum case.
Since $\tilde{r}=r f_R^{1/2}$, the horizon condition becomes
\begin{equation}
1-\frac{2M_0}{r f_R^{1/2}}\hat{M}(z)=0 \ .
\end{equation}
Using the definitions $\gamma\equiv \sqrt{q/\beta}$, $l_\beta^2\equiv 1/(\kappa^2\beta^2)$, $r_q^2\equiv \kappa^2 q^2/(4\pi)$, and $r_S\equiv 2M_0$, the condition $A(r)=0$ becomes
\begin{equation}\label{eq:horizon}
1+2(4\pi)^{3/4}\left(\frac{r_q}{r_S}\right)\sqrt{\frac{r_q}{l_\beta}}G(z)=(4\pi)^{1/4}\left(\frac{r_q}{r_S}\right)\sqrt{\frac{l_\beta}{r_q}}zf_R^{1/2} \ .
\end{equation}
The exterior horizon occurs for $z\gg1$,  where $f_R\approx 1$ and $G(z)\approx 0$. For astrophysical black holes (with low charge to mass ratio, $r_q/r_S\ll1$), it is easy to see that in this limit the left-hand side of (\ref{eq:horizon}) is almost unity, which means that this horizon is located very near the point $z_h\approx 1/\left[(4\pi)^{1/4}\left(\frac{r_q}{r_S}\right)\sqrt{\frac{l_\beta}{r_q}}\right]$. Since $z\equiv r/\gamma$, it follows that $r_h\equiv\gamma z_h\approx r_S$, as expected.

It should be noted that our way of writing (\ref{eq:horizon}) highlights the three scales involved in the problem, namely, the ratio charge-to-mass $r_q/r_S$, the ratio NED-to-charge $l_\beta/r_q$, and the ratio NED-to-Planck given by $\lambda=\kappa^2\beta^2/R_P=l_P^2/l_\beta^2$. 

\subsection{Solving for $\hat{M}(z)$ \label{sec:Mz}}

A closed analytical expression for $\hat{M}(z)$ for arbitrary $\lambda$ is not possible in general. However, a glance at the series expansion of (\ref{eq:Mzt}) near $z\approx 0$ (for finite $\lambda>0$),
\begin{equation}\label{eq:Mzsmall}
\tilde{M}_z\approx \frac{(\pi -2 \lambda )}{16 \pi ^{3/2} \sqrt{\lambda }}z+\frac{ \left(\pi ^2-4 \pi  \lambda +10 \lambda ^2\right)}{32 \pi ^{3/2}  \lambda ^{3/2}}z^3+O(z^5) \ ,
\end{equation}
suggests that the choice $\lambda=\pi/2$ may lead to some simplifications. Luckily this is indeed the case. For that choice of $\lambda$ we can find analytical solutions for that equation. For $\lambda=\pi/2$, Eq.(\ref{eq:Mzt}) becomes
\begin{equation}\label{eq:Mzpihalf}
\tilde{M}_z= \frac{z^3 \left(3+2 z^4\right) \left(1+4 z^4\left(1+z^4\right)-4 z^6\sqrt{1+z^4}\right)}{16 \sqrt{2} \pi  \left(1+z^4\right)^{7/4} \left(1+2 z^4\right)^{3/2}}  \ .
\end{equation}
This expression can be compared with that of GR for the BI model
\begin{equation}\label{eq:MzBI}
\tilde{M}^{GR}_z= \frac{1+z^4-z^2 \sqrt{1+z^4}}{8 \pi  \sqrt{1+z^4}}  \ .
\end{equation}
which is simply $-\frac{z^2}{2\beta^2}$ times Eq.(\ref{tensorBI}), as expected from Eq.(\ref{eq:GR-M}) and the units employed. In the BI case, integrating (\ref{eq:MzBI}) we find that
\begin{eqnarray}
G(z)_{BI}&=& -\frac{(-1)^{\frac{1}{4}} \text{EllipticF}\left[i \text{ArcSinh}\left[(-1)^{\frac{1}{4}} z\right],-1\right]}{12 \pi }\nonumber \\&+&\frac{z\left(\sqrt{1+z^4}-z^2\right)}{24 \pi } +C_{GR} \ ,
\end{eqnarray}
where the integration constant $C_{GR}$ is necessary to get the right asymptotic behavior at $z\to \infty$. Its value is
\begin{eqnarray} \label{CGR}
C_{GR}&=& -\frac{2 (-1)^{1/4} \text{EllipticK}[-1]-(-1)^{3/4} \text{EllipticK}[2]}{12 \pi }\nonumber \\ &=& -\frac{\pi^{1/2}}{24 \Gamma[3/4]^2}   \approx -0.0491809 \ .
\end{eqnarray}
Note that the physical interpretation of this constant is immediate, as it turns out to be $-1/8 \pi$ times the universal energy constant defined below Eq.(\ref{energy}).

In the $f(R)$ case with $\lambda=\pi/2$, the solution is more complicated. In terms of a series expansion, for small $z$ we find
\begin{eqnarray}
G^{z\to 0}_{\frac{\pi}{2}}&=& \frac{ P_1(z)+P_2(z)\text{AF1}\left[\frac{1}{2},\frac{1}{4},-\frac{1}{2},\frac{3}{2},-z^4,-2 z^4\right]}{48 \sqrt{2} \pi  \left(1+3 z^4+2 z^8\right)}\\
P_1(z)&=& \left(1+z^4\right)^{1/4} \sqrt{1+2 z^4}\left[1+4 z^4(1+z^4) \right. \nonumber \\ &-&\left.12 z^2(2+3z^4)\sqrt{1+z^4}\right] \\
P_2(z)&=& 24z^2 \left(1+3 z^4+2 z^{8}\right) \ ,
\end{eqnarray}
where $AF1$ is the Appell hypergeometric function. In order to find a solution for $z\to \infty$, it is convenient to express (\ref{eq:Mzt}) in terms of the variable $x=1/z$. The solution for large $z$ can then be expressed as
\begin{eqnarray}
G^{z\to \infty}_{\frac{\pi}{2}}&=& \frac{ Q_1(z)-Q_2(z)\text{AF1}\left[\frac{1}{4},\frac{-3}{4},\frac{1}{2},\frac{5}{4},\frac{-1}{z^4},\frac{-1}{2 z^4}\right]}{48 \sqrt{2} \pi  z^2 \left(1+3 z^4+2 z^8\right)}\\
Q_1(z)&=& \left(1+z^4\right)^{1/4} \sqrt{1+2 z^4} \left[z^2\left(1+4z^4\left(1+z^4\right)\right) \right. \nonumber \\ &+&\left.4\left(4+6z^4-z^8\right) \sqrt{1+z^4}\right] \\
Q_2(z)&=& 16 \sqrt{2} z \left(1+3 z^4+2 z^8\right)\ .
\end{eqnarray}

These expressions for $G^{z\to 0}_{\frac{\pi}{2}}$ and $G^{z\to \infty}_{\frac{\pi}{2}}$ must be supplemented with an integration constant. In the case of $G^{z\to \infty}_{\frac{\pi}{2}}$, the expansion at $z\to \infty$ leads to $G^{z\to \infty}_{\frac{\pi}{2}}\approx -\frac{1}{16 \pi  z}+\frac{1}{320 \pi  z^5}+O\left[\frac{1}{z}\right]^9$, which recovers the usual GR expression at lowest order and, therefore, does not need the addition of any constant. The integration constant needed by $G^{z\to 0}_{\frac{\pi}{2}}$ can be obtained by forcing the agreement between the $G$'s in the overlapping region. Choosing any point where the two series are well defined, we find that $C_{\frac{\pi}{2}}=G^{z\to \infty}_{\frac{\pi}{2}}(z_0)-G^{z\to 0}_{\frac{\pi}{2}}(z_0)\approx -0.02936$.
If we take, for instance, $z_0=1/\sqrt{2}$, we find
\begin{eqnarray}
C_{\frac{\pi}{2}}&=&\frac{5^{3/4}}{2 \sqrt{6} \pi }-\frac{\sqrt{2} \text{AF1}\left[\frac{1}{4},-\frac{3}{4},\frac{1}{2},\frac{5}{4},-4,-2\right]}{3 \pi }\nonumber \\ &-& \frac{\text{AF1}\left[\frac{1}{2},\frac{1}{4},-\frac{1}{2},\frac{3}{2},-\frac{1}{4},-\frac{1}{2}\right]}{4 \sqrt{2} \pi }\ .
\end{eqnarray}
which fits with the value obtained above. Therefore, near the origin the solution is given by
\begin{equation}
\hat{G}^{z\to 0}_{\frac{\pi}{2}}=C_{\frac{\pi}{2}}+{G}^{z\to 0}_{\frac{\pi}{2}} \ .
\end{equation}

The expansion of $\hat{G}$ near the origin leads to the following expression:
\begin{equation}
\hat{G}^{z\to 0}_{\frac{\pi}{2}}=C_{\frac{\pi}{2}}+\frac{1}{48 \sqrt{2} \pi }+\frac{3 z^4}{64 \sqrt{2} \pi }+\ldots \ ,
\end{equation}
which suggests the definition of a new constant
\begin{equation}
\hat{C}_{\frac{\pi}{2}}=C_{\frac{\pi}{2}}+\frac{1}{48 \sqrt{2} \pi }
\end{equation}
in terms of which $\hat{G}^{z\to 0}_{\frac{\pi}{2}}\approx \hat{C}_{\frac{\pi}{2}}+\frac{3 z^4}{64 \sqrt{2} \pi }+\ldots$. This definition  nicely fits with the expansion corresponding to general $\lambda$, which is of the form
\begin{equation}
{G}^{z\to 0}_{\lambda}=C_{\lambda}+\frac{(\pi -2 \lambda ) z^2}{32 \pi ^{3/2} \sqrt{\lambda }}+\frac{\left(\pi ^2-4 \pi  \lambda +10 \lambda ^2\right) z^4}{128 \pi ^{3/2} \lambda ^{3/2}}+\ldots \
\end{equation}
It should be noted that the constant of integration $C_\lambda$ is negative from $\lambda=0$ to $\lambda\approx 18.6444$.

\begin{figure}
\includegraphics[width=8.6cm,height=5cm]{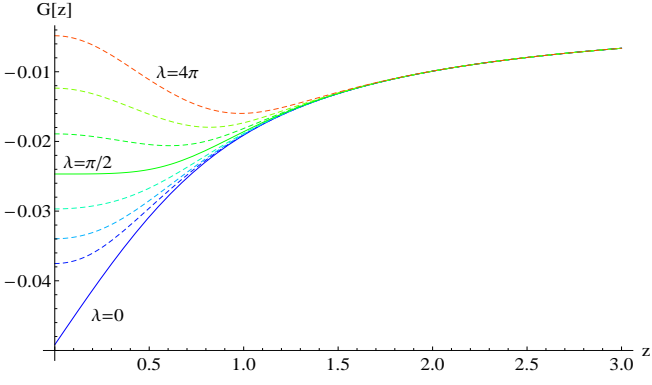}
\caption{From bottom to top representation of the function $G(z)$ for small and medium values of $\lambda=\frac{\pi}{16},\frac{\pi}{8},\frac{\pi}{4}, \pi, 2\pi, 4\pi $ (dashed curves). The solid green curve represents the special case $\lambda=\frac{\pi}{2}$ and the solid blue one gives the GR ($\lambda=0$) case. \label{fig:Gmedium}}
\end{figure}

\begin{figure}
\includegraphics[width=8.6cm,height=5cm]{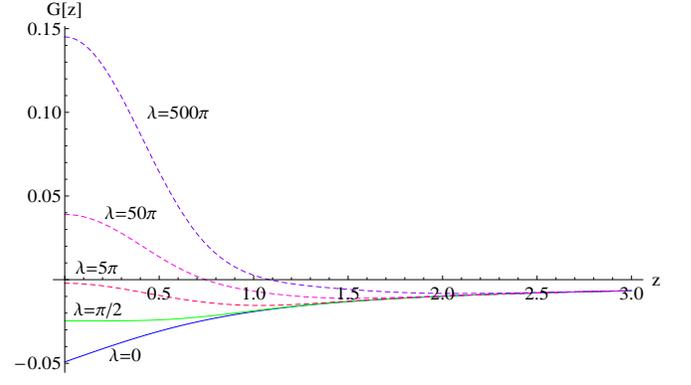}
\caption{From bottom to top representation of the function $G(z)$ for large values of $\lambda=10\times \frac{\pi}{2}, 10^2\times\frac{\pi}{2}, 10^3\times\frac{\pi}{2}$ (dashed curves).  The solid green curve represents the case $\lambda=\frac{\pi}{2}$ and the solid blue one the GR ($\lambda=0$) solution. \label{fig:Glarge}}
\end{figure}

As we will see at once, the $\lambda=\pi/2$ solution is very useful to understand how the Planck length modifies the BI geometry of GR. Since we do not have complete analytical solutions for general $\lambda\neq \pi/2$, each case must be computed separately and, in this sense, the graphical comparison with that solution resulting from a numerical integration will be very illustrative (see Figs.\ref{fig:Gmedium}  and \ref{fig:Glarge}). In this sense, it is important to note that the change in the function $G_\lambda(z)$ observed in Fig.\ref{fig:Gmedium} is smooth as $\lambda$ is continuously increased from zero to positive values. This, however, requires an explanation, because from the expansion (\ref{eq:Mzsmall}) one cannot recover (\ref{eq:MzBI}) in the limit $\lambda\to 0$. In (\ref{eq:Mzsmall}) we considered an expansion for $z\to 0$ with finite $\lambda>0$. If we consider instead an expansion for $\lambda\to 0$ with finite $z>0$, to first order in $\lambda$ we obtain
\begin{equation}\label{eq:Mzlambda}
\tilde{M}_z\approx \tilde{M}_z^{GR}+\frac{\L_1(z)+L_2(z) \sqrt{1+z^4}  }{16 \pi ^2 z^2 \left(1+z^4\right)^{3/2}}\lambda+\ldots \ ,
\end{equation}
where $L_1(z)=z^2-5 z^6-4 z^{10}$ and $L_2(z)=-2+3 z^4 +4 z^8$. This expression is in agreement with (\ref{eq:MzBI}) when $\lambda\to 0$ and is valid as long as the Planck scale corrections can be treated as small perturbations (recall that $\lambda=l_P^2/l_\beta^2$) in the BI background of GR. The perturbative expansion in $\lambda$, however, is not valid when we are near the singularity and the effects of the modified gravitational dynamics are important. Therefore, a faithful description of the geometry for small (but finite) $\lambda$ in the limit $z\to 0$ would require to sum over all the terms in the infinite perturbative series expansion (\ref{eq:Mzlambda}) or, equivalently, consider the non-perturbative expansion about $z\to 0$ given in (\ref{eq:Mzsmall}). This will be clearly seen in the following sections, where we study the properties of the singularity by computing the Kretschmann scalar.

\subsection{Kretschmann scalar}

Given a line element of the form $ds^2=-A(r)dt^2+B(r) dr^2+r^2d\Omega^2$, the Kretschmann scalar $\text{Kret}=R_{\alpha\beta\gamma\delta}R^{\alpha\beta\gamma\delta}$ is given by
\begin{eqnarray}\label{eq:Kret}
\text{Kret}(r)&= &\frac{4 \left(-1+\frac{1}{B}\right)^2}{r^4}+\frac{2 A_r^2}{A^2 B^2 r^2}+\frac{2 B_r^2}{B^4 r^2} \\ &+&\frac{\left(A A_r B_r+B \left(A_r^2-2 A A_{rr}\right)\right)^2}{4 A^4 B^4},\nonumber
\end{eqnarray}
where we are using the notation $A_r \equiv dA/dr$. Given our decomposition of the metric, we have
\begin{eqnarray}\label{eq:A-def}
A(z)&=&\frac{1}{f_R}\left[1-\frac{\left(1+\delta_1 G(z)\right)}{\delta_2 z f_R^{1/2}}\right] \\
B(z) &=& \frac{1}{f_R \left[1-\frac{\left(1+\delta_1 G(z)\right)}{\delta_2 z f_R^{1/2}}\right]}\left(\frac{d(z f_R^{1/2} )}{dz}\right)^2 \ , \label{eq:B-def}
\end{eqnarray}
where the constants $\delta_1$ and $\delta_2$ take the form
\begin{eqnarray}
\delta_1&=& 2(4\pi)^{3/4}\left(\frac{r_q}{r_S}\right)\sqrt{\frac{r_q}{l_\beta}} \\
\delta_2&=& {(4\pi)^{1/4}\left(\frac{r_q}{r_S}\right)\sqrt{\frac{l_\beta}{r_q}}}.
\end{eqnarray}
Since our solutions for the metric are series expansions, there is no chance to write the Kretschmann scalar in a simple form. We will content ourselves with series expansions about the origin to check whether the singular behavior of the metric improves or worsens in these models with respect to the GR solution. Indeed this is enough, as the $f(R)$ theory chosen only deviates from the GR result precisely in that region. We will thus begin by computing the behavior of the BI model in the case of GR.

\subsubsection{Born-Infeld in GR}

In this case, an expansion around the point $z=0$ leads to
\begin{eqnarray}\label{eq:Kret-BI}
\text{Kret}(z)&=& \frac{12 \left(1+C_{GR} \delta _1\right)^2}{\delta _2^2 z^6}+\frac{\delta _1(1+C_{GR} \delta _1)}{\pi  \delta _2^2 z^5}+\nonumber \\ &+& \frac{\delta _1^2}{16 \pi ^2 \delta _2^2 z^4}-\frac{\delta _1^2}{24 \pi ^2 \delta _2^2 z^2}+ \\
&+&\frac{3 \delta _1 \left(1+C_{GR} \delta _1\right)}{10 \pi  \delta _2^2 z}+\frac{13 \delta _1^2}{240 \pi ^2 \delta _2^2}+O[z^1], \nonumber
\end{eqnarray}
where $\text{Kret}(z)=\gamma^4\text{Kret}(r)$, and the constant $C_{GR}$ has been already defined in Eq.(\ref{CGR}).
It is worth noting that for Maxwell electrodynamics the Kretschmann scalar is given by
\begin{equation}\label{eq:Kret-GRM}
\text{Kret}(r)=\frac{48 M_0^2}{r^6}-\frac{48 M_0 r_q^2}{r^7}+\frac{14 r_q^4}{r^8} \ ,
\end{equation}
which diverges as $\sim r^{-8}$ near the origin and indicates that the charge increases the intensity of the divergence. In the BI case, the maximum divergence is of order $\sim r^{-6}$ in general but if the combination of constants $\delta_1$ is constrained to take the value $1+\delta_1 C_{GR}=0$ (because $C_{GR}=C_{\lambda=0}<0$, see Fig.\ref{fig:Gmedium}) then the divergence is softened to the order of $\sim r^{-4}$:
\begin{equation}\label{eq:Kret-BIconstrained}
\text{Kret}(z)=\frac{1}{16\pi ^2 \delta _2^2C_{GR}^2}\left(\frac{1}{z^4}-\frac{2}{3 z^2}+\frac{13}{15 }+\ldots\right)
\end{equation}
Note that, as pointed out at the end of section (\ref{sec:Mz}), the series expansions (\ref{eq:Kret-BI}) and (\ref{eq:Kret-BIconstrained}) are valid for finite values of $l_\beta$, while the limit $l_\beta\to 0$ would require to sum over the infinite terms of the series. This justifies why it is not obvious how to get (\ref{eq:Kret-GRM}) from (\ref{eq:Kret-BI}) taking the limit $l_\beta\to 0$. Similar situations will arise in the $\lambda\neq 0$ cases discussed below. 

The function $A(z)$ near the origin behaves as
\begin{equation}
A(z)\approx -\frac{1+C_{\text{GR}} \delta _1}{z \delta _2}+\left(1-\frac{\delta _1}{8 \pi  \delta _2}\right)+\frac{z^2 \delta _1}{24 \pi  \delta _2}-\frac{z^4 \delta _1}{80 \pi  \delta _2}+\ldots \ .
\end{equation}
With the conditions $1+C_{\text{GR}} \delta _1=0$ and $\delta _2\to -\frac{1}{8 \pi  C_{\text{GR}}}$ this function becomes
\begin{equation} \label{Acenter}
A(z)\approx \frac{z^2}{3}-\frac{z^4}{10}+\frac{z^8}{72} + \ldots
\end{equation}
These special values of $\delta_1$ and $\delta_2$, which are equivalent to constraining the ratios $r_q/r_S$ and $r_q/l_\beta$, receive a nice interpretation. Indeed, using the definitions introduced so far the first condition translates into

\begin{equation}\label{massenergy}
M_0=-8\pi C_{GR}\kappa^2 q^{3/2} \beta^{1/2},
\end{equation}
so taking into account the value of $C_{GR}$ given by (\ref{CGR}) and the expression of the BI field energy (\ref{energy}) this simply means $M_0=\kappa^2 \varepsilon$. On the other hand the second condition above may be written as $\kappa^2 q \beta=4\pi$. These conditions represent limiting values of the BI and the black hole parameters splitting two different regions of black hole configurations (see Refs.\cite{BI-gravity,BI-general}). The first condition splits the metrics into those diverging to $\pm \infty$ at the center, with equation (\ref{massenergy}) being the transition value between both regimens, and for which the metric becomes finite at the center. In the latter case, the addition of the second condition makes the metric to vanish at the center, as shown in Eq.(\ref{Acenter}). When $M_0-\kappa^2\varepsilon(q)<0$ black holes with two horizons (inner and event), extreme black holes or naked singularities may be found (see Fig.{\ref{fig:GRhorizons}), while for $M_0-\kappa^2\varepsilon(q)>0$ a single event horizon appears.

\begin{figure}
\includegraphics[width=8.6cm,height=5cm]{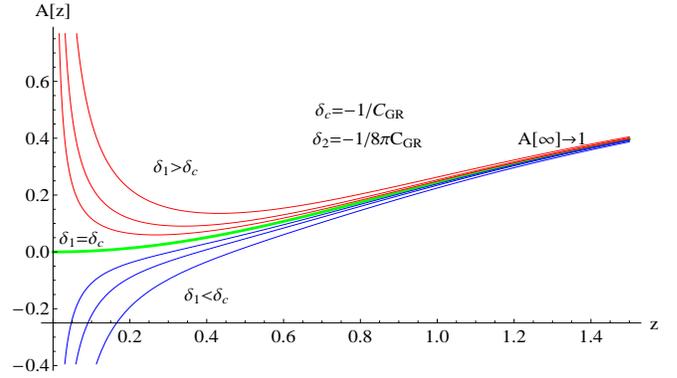}
\caption{Behaviour of the metric function $A(z)$ for GR-BI black holes, according to the sign of $1+C_{GR} \delta_1$ or, alternatively, of $M_0-\kappa^2\varepsilon(q)$. We are plotting the case $\delta_2=\frac{\delta_1}{8\pi}$. Note that the other cases for $M_0=\kappa^2\varepsilon(q)$ may be obtained by moving the associated curve above or below the $A=0$ axis according to whether $\delta_2>\frac{\delta_1}{8\pi}$ or $\delta_2<\frac{\delta_1}{8\pi}$, respectively. \label{fig:GRhorizons}}
\end{figure}

When $1+C_{\text{GR}} \delta _1\neq 0$, the function $B(z)$ near the origin behaves as
\begin{eqnarray}
B(z)&\approx &-\frac{\delta _2 z}{1+C_{\text{GR}} \delta _1}+\frac{\delta _2\left(\delta _1 -8 \pi  \delta _2\right) z^2}{8 \pi  \left(1+C_{\text{GR}} \delta _1\right)^2}\nonumber\\&-&\frac{\delta _2 \left(\delta _1-8 \pi  \delta _2\right)^2 z^3}{64 \pi ^2 \left(1+C_{\text{GR}} \delta _1\right)^3}
+\ldots \ .
\end{eqnarray}
With the conditions $1+C_{\text{GR}} \delta _1=0$ and $\delta _2\to -\frac{1}{8 \pi  C_{\text{GR}}}$ this function becomes
\begin{equation}
B(z)\approx\frac{3}{z^2}+\frac{9}{10}+\frac{27 z^2}{100}-\frac{11 z^4}{250}-\frac{507 z^6}{10000}+O[z]^7.
\end{equation}
Note that to obtain this last result one must impose the conditions on $\delta_1$ and $\delta_2$ before expanding (\ref{eq:B-def}) about $z\approx 0$.

\subsubsection{Born-Infeld for $\lambda\neq 0,\pi/2$}

In this case the analytical expansion leads to
\begin{equation}
\text{Kret}(z)= \frac{a}{z^{12}}+\frac{b}{z^{10}}+\frac{c}{z^{8}}+\frac{d}{z^{6}}+\frac{e}{z^{4}}+\frac{f}{z^{2}}+a_0+\ldots
\end{equation}
The coefficients are functions of $\delta_1$, $\delta_2$, $\lambda$, and the integration constant $C_\lambda$  necessary to match the $z\to \infty$ and the $z\to 0$ series expansions. One can choose various combinations of those parameters that make some of the coefficients in the series expansion vanish. In particular, the choice
\begin{eqnarray}
\delta_2&=&\frac{\sqrt{\pi }(1+C_\lambda \delta_1)}{\sqrt{\lambda }} \\
\delta_1&=&\frac{16 \pi ^{3/2}}{\sqrt{\lambda }-16 C_\lambda \pi ^{3/2}},
\end{eqnarray}
leads to the largest simplification
\begin{eqnarray}
\text{Kret}(z)&=&\frac{79}{4 z^4}+\frac{77 \left(3 \pi ^2-20 \pi  \lambda +10 \lambda ^2\right)}{12 (\pi -2 \lambda ) \lambda  z^2} \nonumber \\&+&\frac{-243 \pi ^4-3600  \pi ^3 \lambda +23300 \pi ^2 \lambda ^2}{96 (\pi -2 \lambda )^2 \lambda ^2}  \\ &+& \frac{-8960 \pi  \lambda ^3+30740 \lambda ^4}{96 (\pi -2 \lambda )^2 \lambda ^2}+\ldots \nonumber
\end{eqnarray}

The function $A(z)$ near the origin behaves as
\begin{eqnarray}
A(z)&\approx& \frac{\pi  z^2 \left(-\sqrt{\pi } \left(1+C_{\lambda } \delta _1\right)+\sqrt{\lambda } \delta _2\right)}{\lambda ^{3/2} \delta _2}
\\&-&\frac{z^4 (\pi -2 \lambda ) (-48 \pi ^{3/2} (1+C_{\lambda } \delta _1)+\sqrt{\lambda } (\delta _1+32 \pi  \delta _2))}{32 \lambda ^{5/2} \delta _2}. \nonumber
\end{eqnarray}
With the conditions on $\delta_2$ and $\delta_1$ discussed above, this function becomes
\begin{eqnarray}
A(z)&\approx& -\frac{\pi  (\pi -2 \lambda )^2z^6}{4 \lambda ^3}\\&+&\frac{\pi  \left(15 \pi ^3-82 \pi ^2 \lambda +130 \pi  \lambda ^2-52 \lambda ^3\right) z^8}{24 \lambda ^4} + \ldots \nonumber
\end{eqnarray}

The function $B(z)$ near the origin behaves as
\begin{eqnarray}
B(z)&\approx &\frac{\delta_2 (\pi -2 \lambda )^2 z^4}{\left(-\sqrt{\pi } (1+C_\lambda\delta_1)+\sqrt{\lambda }\delta_2 \right) \lambda ^{3/2}}\\ &-& \frac{a_\lambda z^6}{32 \pi  \left(-\sqrt{\pi } (1+C_\lambda \delta_1)+\sqrt{\lambda }\delta_2\right)^2 \lambda ^{5/2}}+\ldots \nonumber
\end{eqnarray}
where $a_{\lambda}$ is an involved function of $\lambda$, $C$, $\delta_1$ and $\delta_2$. Again, with the conditions on $\delta_2$ and $\delta_1$ discussed above, this function becomes
\begin{equation}
B(z)\approx -4+\frac{\left(6 \pi ^2+20\lambda (\lambda -2 \pi )\right) z^2}{3 \pi  \lambda -6 \lambda ^2}+O(z^4)
\end{equation}

\subsubsection{Born-Infeld in $\lambda=\pi/2$}

In this case, the strongest divergence of the Kretschmann goes as $\sim z^{-20}+O(z^{-16})+\ldots$. If we impose the following condition
\begin{equation}
\delta_2=\sqrt{2}(1+\hat{C}_{\frac{\pi}{2}}\delta_1) \ ,
\end{equation}
then the leading order divergences are suppressed to $\sim z^{-12}$. Demanding that the coefficient of that term also vanishes, we find the condition
\begin{equation}
\delta_1=  \frac{32 \pi }{\sqrt{2}-32 \pi \hat{C}_{\frac{\pi}{2}} } \ ,
\end{equation}
which leads to
\begin{eqnarray}
\text{Kret}(z)&=&\frac{739}{64 z^4}-\frac{797}{15 z^2}+\frac{290787}{3200}+\ldots
\end{eqnarray}
This result indicates that the modified gravitational action cannot soften the strength of the divergence below the limits of the standard BI model in GR. 

The function $A(z)$ near the origin behaves as
\begin{eqnarray}
A(z)&\approx& -\frac{2 \left(\sqrt{2}(1+\hat{C}_{\frac{\pi}{2}} \delta _1)-\delta _2\right) z^2}{\delta _2}\\&-&\frac{3 \left[\delta _1-16 \pi  \left(3 \sqrt{2}(1+\hat{C}_{\frac{\pi}{2}} \delta _1)-2 \delta _2\right)\right] z^6}{32 \pi  \delta _2}+\ldots \nonumber
\end{eqnarray}
With the conditions on $\delta_2$ and $\delta_1$ discussed above, this function becomes
\begin{equation}
A(z)\approx -\frac{9 z^{10}}{8}+O(z^{11}).
\end{equation}

The function $B(z)$ near the origin behaves as
\begin{equation}
B(z)\approx -\frac{9 \delta _2 z^8}{\sqrt{2}(1+\hat{C}_{\frac{\pi}{2}} \delta _1)-\delta _2}+O(z^{11}).
\end{equation}
With the conditions on $\delta_2$ and $\delta_1$ discussed above, this function becomes
\begin{equation}
B(z)\approx -16-\frac{512 z^2}{15}-\frac{7084 z^4}{225}-\frac{598016 z^6}{23625}-\ldots.
\end{equation}

\subsection{Inner horizons}

Given the units that we are using, for astrophysical black holes with low charge to mass ratio ($\delta_1\ll 1$), the existence and location of the outer horizon will be only slightly modified as compared with the GR case. For this reason, in this work we focus on the analysis of the inner horizons. The effects of the Planck-scale modified dynamics on microscopic black holes, whose outer horizon may be significantly affected by the internal structure near the singularity, will be explored elsewhere.

Horizons are determined by the cuts between the functions $1+\delta_1 G(z)$ and $\delta_2 z f_R^{1/2}$. In Fig.\ref{fig:horizons+} we can see how these curves change as we modify $\lambda$ and $l_\beta$. For the GR case ($\lambda=0$), for instance, we see that changing $l_\beta$ may lead to the existence or absence of one inner horizon. To have an inner horizon we need a small enough value of $l_\beta$. The structure of $\lambda=0$ is smoothly deformed as $\lambda$ is increased. As depicted in Fig.\ref{fig:horizons+}, the case $\lambda=\pi/2$ also has an horizon for small $l_\beta$, though it disappears for larger values. For values of $\lambda$ larger than $\pi/2$, the function $G(z)$ develops a minimum and a maximum near the origin, as we saw in Figs.\ref{fig:Gmedium} and \ref{fig:Glarge}, which may lead to the formation of up to two inner horizons for some combinations of pairs $(\lambda, l_\beta)$. In this section we describe the structure of inner horizons for the various combinations of such pairs.

\begin{figure}
\includegraphics[width=8.6cm,height=5cm]{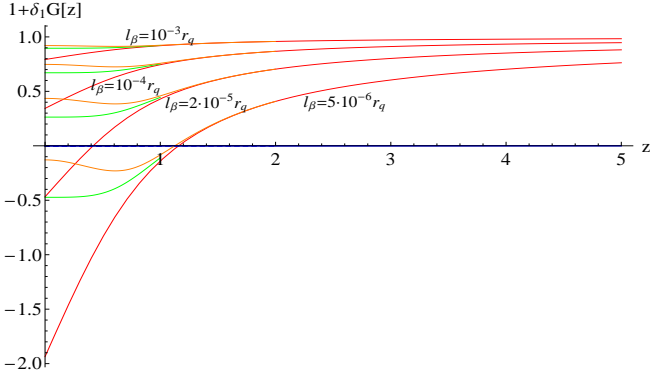}
\caption{Representation of the curves $1+\delta_1 G(z)$ (various colors) and $\delta_2 z f_R^{1/2}$ (blue) for the GR case (red), the $\lambda=\pi/2$ case (green) and the $\lambda=\pi$ case (orange) for the relations $r_q=10^{-2}r_S$ and $l_\beta=10^{-3}r_q,10^{-4}r_q, 2\times 10^{-5}r_q,5\times 10^{-6}r_q$. \label{fig:horizons+}}
\end{figure}

\subsubsection{Critical value for $l_\beta$}

Let us define $l^{crit}_\beta$ as the value of $l_\beta$ above which there is no inner horizon. A natural way to determine the value of $l^{crit}_\beta$ is to look at the GR solution. In this case, the inner horizon disappears when the function $\hat{M}(z)\equiv 1+\delta_1 G(z)$ vanishes at the origin. The value of $l_\beta$ that fulfills this condition turns out to satisfy the relation $1+\delta_1 C_{GR}=0$, being $C_{GR}=G(0)$ (the integration constant needed to match with the $z\gg 1$ solution). This constraint gives $l^{crit}_\beta$ in terms of $r_q$ and $r_S$ as follows
\begin{equation}
l^{crit}_\beta= \frac{32\pi^{3/2} C^2 r_q^3}{r_S^2}.
\end{equation}
Remarkably, this relation not only holds for the GR case with $C=C_{GR}$ but it does also hold for $C=\hat{C}_{\frac{\pi}{2}}$ and the general case $C=C_\lambda$ with $\lambda<\pi/2$ (see Fig.\ref{fig:lbetam}).

\begin{figure}
\includegraphics[width=8.6cm,height=5cm]{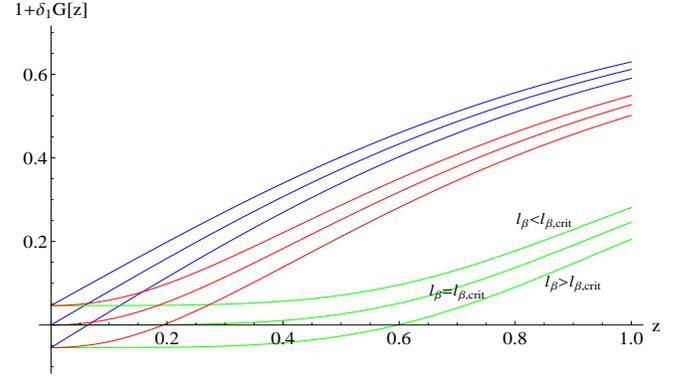}
\caption{Representation of the curve $1+\delta_1 G(z)$ for $l_\beta=1.1\times l_\beta^{crit}$ , $l_\beta=l_\beta^{crit}$ , and $l_\beta=0.9\times l_\beta^{crit}$ (lower, central, and upper curves respectively for each color) for $\lambda=0$ (blue), $\lambda=\pi/2$ (green), and $\lambda=1/10$ (red). Note that $l_\beta^{crit}(0)\neq l_\beta^{crit}(\pi/2)\neq l_\beta^{crit}(1/10)$. \label{fig:lbetam}}
\end{figure}

\begin{figure}
\includegraphics[width=8.6cm,height=5cm]{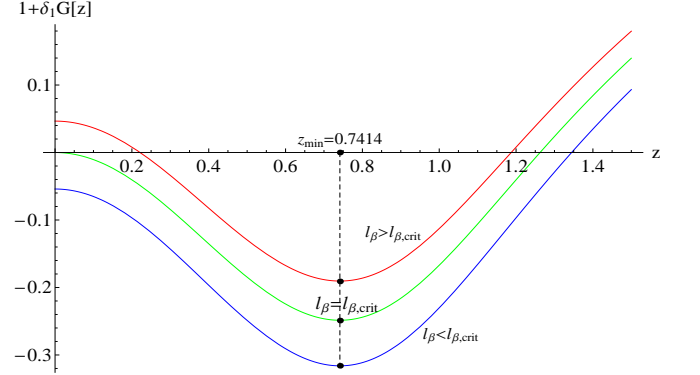}
\caption{Representation of the curve $1+\delta_1 G(z)$ for $l_\beta=1.1\times l_\beta^{crit}$ (red), $l_\beta=l_\beta^{crit}$ (green), and $l_\beta=0.9\times l_\beta^{crit}$ (orange) for $\lambda=3\pi/2$. Note that the minima occur at the same location $z_{min}\simeq 0.7414$, where $dG/dz=0$. \label{fig:lbetap}}
\end{figure}

The structure corresponding to $\lambda<\pi/2$ is standard since only one inner horizon may arise if any. For $\lambda>\pi/2$ one can also find up to two inner horizons. The critical length $l_\beta^{crit}$ in this case sets the value of $l_\beta$ that first shows two inner horizons, being the innermost one located at $z=0$ (see Fig.\ref{fig:lbetap}). For larger values of $l_\beta$, the innermost horizon moves away from the origin and approaches the second inner horizon. Eventually, the two inner horizons coincide and for larger values of $l_\beta$ no inner horizon arises. This behaviour with $\lambda$ is explained by the fact that for $\lambda<\pi/2$ the slope of the function $1+ \delta_1 G(z)$ near the center is positive, while when $\lambda>\pi/2$ it becomes negative, thus allowing up to two cuts with the curve $\delta_2 z f_R^{1/2}$ when $l_{\beta}>l_{\beta}^{crit}$. In this sense, the special case $\lambda=\pi/2$ represents the transition value between these regions and corresponding to a vanishing slope of the curve $1+\delta_1 G(z)$ at the center, while $\lambda=0$ (the GR case) becomes an upper limit for this slope. Note that the value of the function $1+\delta_1 G(0)$ does not depend on $\lambda$, but just on the value of $l_{\beta}$ as compared to the one of $l_{\beta}^{crit}$.

\subsubsection{Critical value for $\delta_1$}

The previous discussion can be reinterpreted in a different way. First we should note that our definition of $G(z)$ is missing its dependence on the parameter $\lambda$ and, therefore, we should use the alternative notation $G(z;\lambda)$. This function $G(z;\lambda)$ can be expanded near the origin as $G(z;\lambda)=C_\lambda+a_\lambda z^2+b_\lambda z^4+\ldots$. Therefore, the function $\hat{M}(z;\delta_1)=1+\delta_1 G(z;\lambda)$ at the origin takes the value $\hat{M}(0;\delta_1)=1+\delta_1 C_\lambda$. For those values of $\lambda$ for which $C_\lambda<0$, the special condition $\hat{M}(0;\delta_1)=0$ picks out a special value of $\delta_1$, which we may denote as $\delta_1^\lambda=-1/C_\lambda$. With this notation, we have $G(z;\lambda)=-\frac{1}{\delta_1^\lambda}+{a}_\lambda z^2+{b}_\lambda z^4+\ldots$ , which explains why those configurations with $\delta_1>\delta_1^\lambda$ have $\hat{M}(0;\delta_1)\equiv 1-\delta_1/\delta_1^\lambda<0$ and those with $\delta_1<\delta_1^\lambda$ have $\hat{M}(0;\delta_1)>0$. This notation also makes clear why the plots presented in Fig.\ref{fig:lbetam} for curves with different $\lambda$ but the same $\delta_1$ hit the vertical axis at the same point. 

\subsubsection{Characterization of double inner horizons}

We have just seen that if $\hat{M}(0;\delta_1)<0$ then we must have an inner horizon. However, $\hat{M}(0)>0$ does not necessarily mean that we do not have an inner horizon. In this situation we may have none (if $\lambda<\pi/2$) or two (for some values of $\delta_1$ if $\lambda>\pi/2$) or even one degenerate (extreme). Now we will detail the conditions for the existence of two inner horizons for a given $\lambda$.

Consider Fig.\ref{fig:lbetap}. The minimum of those curves are located at the same position $z_{min}^\lambda$. The functional dependence of $z_{min}^\lambda$ with $\lambda$ can be obtained analytically but the resulting expression is too long to be written here and not too much illuminating anyhow. Nonetheless, what really matters is that we know exactly where $z_{min}^\lambda$ is located for a given $\lambda$. The next step is to study the evolution of $\hat{M}(z_{min}^\lambda;\delta_1)=1+\delta_1 G(z_{min}^\lambda;\lambda)$ as $\delta_1$ is changed (see Fig. \ref{fig:MzminEvol}). 

\begin{figure}
\includegraphics[width=8.6cm,height=5cm]{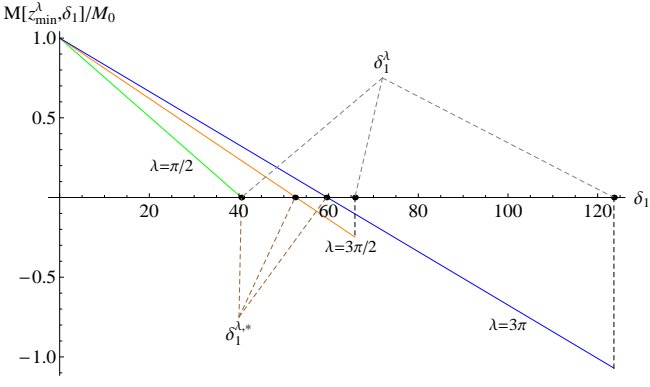}
\caption{Evolution of $\hat{M}(z_{min}^\lambda;\delta_1)$ as $\delta_1$ is changed for $\lambda=\pi/2, 3\pi/2, 3\pi$ (green, orange, and blue respectively). The $x$-axis represents the variable $\delta_1$, which goes from $\delta_1=0$ to $\delta_1^\lambda$ in each case. Note that for $\lambda=\pi/2$ we have $\delta_1^\lambda=\delta_1^{\lambda,*}$. \label{fig:MzminEvol}}
\end{figure}

When the curve $\hat{M}(z_{min}^\lambda;\delta_1)$ cuts the horizontal axis, $\hat{M}(z_{min}^\lambda,\delta_1^{\lambda,*})=0$, then that value of $\delta_1^{\lambda,*}$ represents a degenerate double horizon. In other words, for $\delta_1=\delta_1^{\lambda}$ we have two inner horizons, being the innermost located at $z=0$, where $\hat{M}(0;\delta_1^\lambda)=0$; for $\delta_1^{\lambda,*}<\delta_1<\delta_1^\lambda$ we have two inner horizons;  for $\delta_1=\delta_1^{\lambda,*}$ the two inner horizons converge at the same point; and for $\delta_1<\delta_1^{\lambda,*}$ there are no inner horizons. The curves appearing in Fig.\ref{fig:MzminEvol} are segments of straight lines with an end located at $(\delta_1^\lambda,\hat{M}(z_{min}^\lambda;\delta_1^\lambda),\delta_1^\lambda)$ and the other at $(0,1)$. This means that those lines can be expressed as follows
\begin{equation}
y=1-\left(\frac{1-\hat{M}(z_{min}^\lambda;\delta_1^\lambda)}{\delta_1^\lambda}\right) \delta_1 \ .
\end{equation}
From this expression it is clear that $\delta_1^{\lambda,*}$ is defined by $y=0$, which leads to $\delta_1^{\lambda,*}=\frac{\delta_1^\lambda}{1-\hat{M}(z_{min}^\lambda;\delta_1^\lambda)}$. Note that both $\delta_1^\lambda$ and $\hat{M}(z_{min}^\lambda;\delta_1^\lambda)$ can be computed numerically for each $\lambda$.

\section{Study of $f(R)=R-R^2/R_P$ \label{R-R2}}

In this case the equation (\ref{eq:Mz}) can be manipulated to allow for a more transparent interpretation and analysis. Taking into account the following identity
\begin{equation}
\frac{1}{f_R^{3/2}}\left(f_R+\frac{z}{2}f_{R,z}\right)=-z^2\frac{d}{dz}\left(\frac{1}{z f_R^{1/2}}\right) \ ,
\end{equation}
we can express (\ref{eq:Mz}) as follows
\begin{equation}\label{eq:Mz2}
\tilde{M}_z=-\frac{z^4}{4}\left(\tilde{f}+\frac{\tilde{\varphi}}{4\pi}\right)\frac{d}{dz}\left(\frac{1}{z f_R^{1/2}}\right) \ .
\end{equation}
This expression can also be written as
\begin{equation}\label{eq:Mz2}
\frac{d}{dz}\left[\tilde{M}+\frac{z^3\left(\tilde{f}+\frac{\tilde{\varphi}}{4\pi}\right)}{4f_R^{1/2}}\right]=\frac{1}{4zf_R^{1/2}}\frac{d}{dz}\left[z^4\left(\tilde{f}+\frac{\tilde{\varphi}}{4\pi}\right)\right] \ .
\end{equation}
The advantage of this expression is not clear yet. However, for the model $f(R)=R-R^2/R_P$ in which $f_R$ vanishes at some small $z$, we will see that the right-hand side diverges as $1/\sqrt{z-z_+}$ (where $z_{+}$ is defined below) and, therefore, its integral is finite. This means that the divergence of $\tilde{M}(z)$ is totally described by the second term on the left-hand side. We thus have fully under control the divergent piece of the mass function. 

We now determine the series expansion of the right-hand side of (\ref{eq:Mz2}) near the divergence. The divergence occurs when $f_R=0$, i.e., at a curvature $R=R_P/2$. For this reason, we write this function as follows ($\tilde{\lambda}\equiv \lambda/\pi$)
\begin{eqnarray}
f_R&=& \frac{\tilde{\lambda}}{\sigma}(\sigma_+-\sigma)(\sigma-\sigma_-) \\
\sigma &=& \sqrt{1+\frac{1}{z^4}} \\
\sigma_{\pm} &=& \sqrt{1+\frac{1}{z^4_{\pm}}} \\
\frac{1}{z^4_{\pm}}&=& \frac{\sqrt{1+4\tilde{\lambda}}}{2\tilde{\lambda}^2}\left(\sqrt{1+4\tilde{\lambda}}\pm(1+2\tilde{\lambda})\right) .
\end{eqnarray}
With this notation, it is easy to see that $f_R$ vanishes at $z=z_+$. For $\tilde{\lambda}\ll 1$ it is also easy to see that $z_+\approx \tilde{\lambda}^{1/2}$, where $\pi\tilde{\lambda}\equiv \kappa^2\beta^2/R_P=l_P^2/l_\beta^2$ and, therefore, $z_+\sim l_P/l_\beta$.

\begin{figure}
\includegraphics[width=8.6cm,height=5cm]{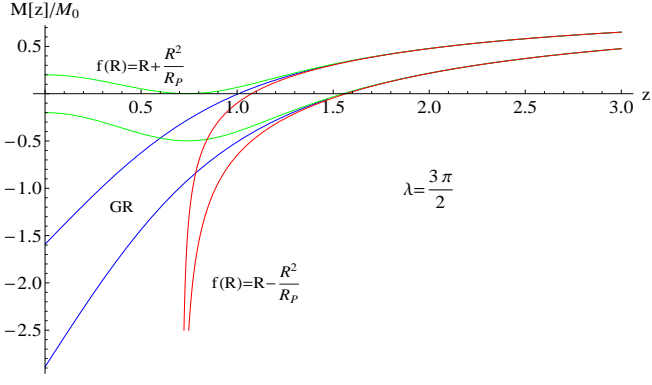}
\caption{Representation of $\hat{M}(z)$ for GR (blue), and $\lambda=3\pi/2$ for $f(R)=R+R^2/R_P$ (green) and $f(R)=R-R^2/R_P$ (red). The curves represent the cases $\delta_1=\delta_1^{\lambda=3\pi/2}(1\pm 2\times 10^{-2})$. Note that GR and $f(R)=R-R^2/R_P$ only present one horizon, whereas $f(R)=R+R^2/R_P$ may develop up to two horizons.  \label{fig:M(z)1}}
\end{figure}

\begin{figure}
\includegraphics[width=8.6cm,height=5cm]{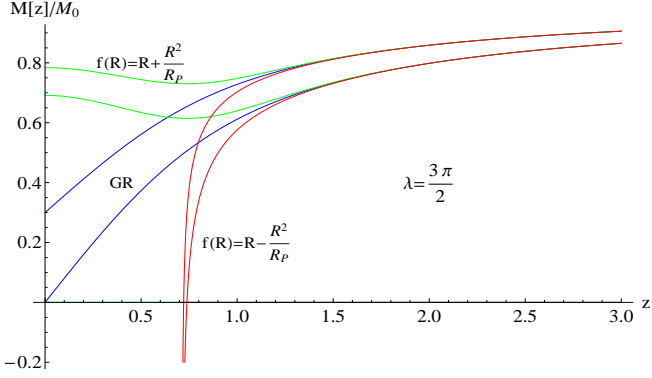}
\caption{Representation of $\hat{M}(z)$ for GR (blue), and $\lambda=3\pi/2$ for $f(R)=R+R^2/R_P$ (green) and $f(R)=R-R^2/R_P$ (red). The curves represent the cases $\delta_1=\delta_1^{\lambda=0}$ and $\delta_1=\delta_1^{\lambda=0}(1- 3\times 10^{-2})$. In these cases the GR solution $\delta_1=\delta_1^{\lambda=0}$ has the inner horizon at $z=0$ and $\delta_1<\delta_1^{\lambda=0}$  has no inner horizon. The case  $f(R)=R-R^2/R_P$  develops an inner horizon as long as $\delta_1>0$. \label{fig:M(z)2}}
\end{figure}

Note that $\sigma_-=1/\sigma_+$, and that as $\tilde{\lambda}\to \infty$ we find $\sigma_+=1^+$ whereas $\sigma_-=1^-$. This means that $\sigma^+>1$ always. It will also be useful to express $\tilde{\lambda}$ in terms of $z_+$ as follows
\begin{equation}
\tilde{\lambda}=\sqrt{z_+^4(1+z_+^4)}\left[1+2z_+^4+2\sqrt{z_+^4(1+z_+^4)}\right].
\end{equation}
Using this expression, we can expand the right-hand side of (\ref{eq:Mz2}) near $z=z_++\epsilon$ as follows
\begin{eqnarray}\label{eq:RHS}
RHS&\approx& \frac{\epsilon^{-1/2}}{4 \pi  \left(1+z_+^4\right) \sqrt{\frac{2}{z_+}+4 z_+^3+4 z_+ \sqrt{1+z_+^4}} }\\&+&\frac{\left(3-23 z_+^4-16 z_+^2 \sqrt{1+z_+^4}\right) \epsilon^{1/2}}{16 \sqrt{2} \pi  \left(1+z_+^4\right)^2 \sqrt{z_++2 z_+^5+2 z_+^3 \sqrt{1+z_+^4}}}\nonumber
\end{eqnarray}
where $\epsilon\equiv \sigma_+-\sigma$. This shows that the integral of this function is finite, as claimed before. The function $\tilde{M}(z)$ is plotted in Figs.\ref{fig:M(z)1} and \ref{fig:M(z)2}. As we see there, an inner horizon always exists, regardless of the model parameters when the constraint $\delta_1>0$ is satisfied.

From this analysis we conclude that for the model $f(R)=R-R^2/R_P$ the mass function diverges as $M(z)\sim 1/f_R^{1/2}\sim 1/\sqrt{z-z_+}$ as we approach $z^+$. The internal region always has an inner horizon as long as $\delta_1>0$ (which is the only physically meaningful situation).

Given the expressions (\ref{eq:Mz2}) and (\ref{eq:RHS}), the structure of the solution near the singular point $z=z_+$ is completely known up to an integration constant coming from the integration of (\ref{eq:RHS}).

\subsection{Metric components}

Using the same decomposition of the mass function $\hat{M}=1+\delta_1 G(z;\lambda)$ as in the $f(R)=R+R^2/R_P$ case, near the critical point $z_+$ the metric functions $A(z)$ and $B(z)$ can be approximated as follows
\begin{eqnarray}
A(z)&=&\frac{a_1}{\epsilon^2}+\frac{a_2}{\epsilon^{3/2}}+\frac{a_3}{\epsilon}+\frac{a_4}{\epsilon^{1/2}}+a_5+a_6 \epsilon^{1/2}+\ldots \\
B(z)&=&\frac{b_1}{\epsilon}+\frac{b_2}{\epsilon^{1/2}}+b_3+b_4 \epsilon^{1/2}+\ldots
\end{eqnarray}
The form of the coefficients can be obtained recursively, being the leading terms given by
\begin{eqnarray}
a_1&=& \frac{\left(-1+\sigma _+\right){}^{5/2} \sigma _+ \delta _1}{16 \pi  \left(1+\sigma _+\right){}^{3/2} \delta _2}\\
b_1&=&\frac{16 \pi  \left(-1+\sigma _+\right) \left(1+\sigma _+\right){}^3 \delta _2}{\sigma _+^2 \sqrt{-1+\sigma _+^2} \delta _1} \ .
\end{eqnarray}

\subsection{Kretschmann scalar}
To study the degree of divergence of the Kretschmann scalar, we will consider the leading order contribution near $z_+$, i.e., when $\epsilon\to 0$,  of the functions $A(z)$ and $B(z)$. For clarity, it is useful to replace the infinitesimal $\epsilon=\sigma_+-\sigma$ by its corresponding expression in terms of $\zeta=z-z_+$. The correspondence can be easily found and is given by
\begin{equation}
\epsilon=-\frac{2\zeta}{z_+^5 \sigma_+}=-\frac{2(\sigma^2_+-1)^{5/4}}{\sigma_+}(z-z_+) \ .
\end{equation}
We thus have the following leading-order expressions as $z\to z_+$
\begin{eqnarray}
A(z)&\approx &\frac{\tilde{a}_1}{(z-z_+)^2}++\ldots \\
B(z)&\approx &\frac{\tilde{b}_1}{z-z_+}+\ldots
\end{eqnarray}
Inserting these expressions into (\ref{eq:Kret}), we find
\begin{equation}
Kret(z)\approx \frac{9 \left(-1+\sigma _+\right) \sigma _+^2 \sqrt{-1+\sigma _+^2} \delta _1^2}{64 \pi ^2\left(1+\sigma _+\right)^3 \delta _2^2}\frac{1}{\left(z-z_+\right)^2}+\ldots
\end{equation}
This result is extremely interesting. It means that even though the metric components are strongly divergent as $z_+$ is approached, the curvature at that point diverges just as $\sim1/(z-z_+)^2$, which is milder than in the case of GR and $f(R)=R+R^2/R_P$ where it goes as $1/z^4$. This sharp difference in the intensity of the divergence is entirely due to the shift of the singularity from $z=0$ to $z=z_+$, because in this case terms such as $(-1+B)^2/z^4$ are no longer divergent. The divergent piece is just due to the last term of (\ref{eq:Kret}).

\section{Conclusions}

We have studied electrically charged black holes within the context of Palatini $f(R)=R\pm R^2/R_P$ gravities, by considering Born-Infeld electrodynamics. This choice of nonlinear electrodynamics as the matter source was chosen in order to break the tracelessness condition of the energy-momentum matter tensor, thus allowing to obtain solutions different from those of GR. In this work we have focused, in particular, in black holes whose charge to mass ratio is small ($r_q/r_S\ll 1$). These black holes only deviate from their GR counterparts near the center, and lead to several relevant modifications: i) new horizons may arise: in the $f(R)=R+R^2/R_P$ case there may be up to two inner horizons (which have been fully characterized), having also the possibility of being degenerate (extreme), ii) in the $f(R)=R-R^2/R_P$ case the corresponding black holes show a similar structure in terms of horizons to that of the Reissner-Nordstr\"om solution of Einstein-Maxwell equations in GR, but the radial coordinate cannot be extended below a minimum value $r_+$,  iii) the strength of the central curvature singularity may be softened due to Palatini $f(R)$ gravity: from $r^{-8}$ of the Reissner-Nordstr\"om solution, and the $r^{-6}$ of BI in GR (or $r^{-4}$ if the combination of the model and black hole parameters are properly chosen), we go to the $r^{-4}$ in Palatini $f(R)=R+R^2/R_P$ gravity but to $(r-r_+)^{-2}$ in Palatini $f(R)=R-R^2/R_P$ gravity.

Our results show that Palatini $f(R)$ theories are able to severely modify the internal structure of electrically charged black holes. In the $f(R)=R-R^2/R_P$ case, the softening of the divergence by the shift of the central singularity to a finite point motivates the study of extensions of Palatini $f(R)$ theories and also other NED models to better understand how the interplay between modified gravity and modified matter sources at very high curvatures may affect this internal black hole structure and ameliorate its singularities, an issue to be explored in future works.

\section*{Acknowledgments}

We are indebted to J. Navarro-Salas for stimulating discussions. G. J. O. has been supported by the Spanish grant FIS2008-06078-C03-02 and the Consolider Programme CPAN (CSD2007-00042). D. R.-G. would like to thank for all their hospitality to the Departamento de F\'isica Te\'orica of Valencia U., where this work initiated during a visit.

\end{document}